\documentclass[a4paper]{article}

\usepackage[english]{babel}
\usepackage[utf8x]{inputenc}
\usepackage[T1]{fontenc}

\usepackage[a4paper,top=3cm,bottom=2cm,left=3cm,right=3cm,marginparwidth=1.75cm]{geometry}

\usepackage{amsmath}
\usepackage{booktabs}
\usepackage{makecell}
\usepackage{multirow}
\usepackage{float}
\usepackage{subfigure}
\usepackage{wrapfig}
\usepackage{caption}
\usepackage{graphicx}
\usepackage[colorinlistoftodos]{todonotes}
\usepackage[colorlinks=true, allcolors=blue]{hyperref}
\newcommand{\ackname}{Acknowledgements}
\usepackage{url}

\title{Deep Convolutional Neural Networks for Noise Detection in ECGs}
\author{
John, Jennifer\\
Stanford Online High School\\
\texttt{jnjohn@ohs.stanford.edu}
\and
Galloway, Conner\\
AliveCor, Inc.\\
\texttt{conner@alivecor.com}
\and
Valys, Alexander\\
AliveCor, Inc.\\
\texttt{avalys@alivecor.com}
}

\begin{document}
\maketitle

\begin{abstract}
Mobile electrocardiogram (ECG) recording technologies represent a promising tool to fight the ongoing epidemic of cardiovascular diseases, which are responsible for more deaths globally than any other cause. While the ability to monitor one’s heart activity at any time in any place is a crucial advantage of such technologies, it is also the cause of a drawback: signal noise due to environmental factors can render the ECGs illegible. In this work, we develop convolutional neural networks (CNNs) to automatically label ECGs for noise, training them on a novel noise-annotated dataset. By reducing distraction from noisy intervals of signals, such networks have the potential to increase the accuracy of models for the detection of atrial fibrillation, long QT syndrome, and other cardiovascular conditions. Comparing several architectures, we find that a 16-layer CNN adapted from the VGG16 network which generates one prediction per second on a 10-second input performs exceptionally well on this task, with an AUC of 0.977.
\end{abstract}

\section{Introduction}

Cardiovascular disease (CVD) is the leading cause of death in the world, claiming more than 17.3 million lives each year \cite{members2017heart}. In order to stop this epidemic, personalized, proactive, and accessible heart care will be essential. This is particularly true given that low- and middle-income countries, where health care is often inaccessible, have disproportionately high rates of deaths from CVD \cite{yusuf2014cardiovascular}.

Mobile electrocardiogram (ECG) recording technologies empower patients to monitor their own heart health by reducing the financial and temporal costs typically associated with recording and analyzing traditional 12-lead ECGs in a hospital. Such technologies have proven capable of diagnosing conditions including atrial fibrillation and long QT syndrome as accurately as a cardiologist. However, noise and signal artifact in ECG signals can restrict performance. Without the additional precision afforded by 12 leads and the standardization of a hospital setting, environmental factors can interfere with a signal, sometimes rendering it impossible to analyze. Movement of the patient while recording, radio frequency interference, and inadequate moisture on recording surfaces can all lead to noise.

In this paper, we develop a model capable of locating noise in ECGs with high accuracy. When coupled with algorithms to detect arrhythmias in ECGs, this model will enable augmented performance by suggesting intervals of a signal to exclude from analysis. With only the relatively clean intervals included, a diagnosis algorithm will be able to focus on the meaningful, if relatively small, features that may be obscured by noise, such as P-waves. Excluding noise also prevents the algorithm from erroneously interpreting features introduced by noise or signal artifact as physiological conditions. For example, some types of noise resemble the sudden deflections seen in premature ventricular contractions (PVCs). Conversely, it is equally important that abnormal morphologies or arrhythmias are not falsely interpreted as noise.

We apply convolutional neural networks (CNNs) to this task, experimenting with several architectures. Based on the assumption of a 3-dimensional volume \cite{CS231n}, CNNs were initially developed for image recognition, winning the 2012 ILSVRC competition \cite{NIPS2012_4824}. However, they have displayed strong performance on a diverse array of tasks, from natural language processing \cite{NIPS2014_5550,DBLP:journals/corr/Kim14f} to video classification \cite{Karpathy_2014_CVPR}. Most relevant to this work, CNNs have been applied to ECG classification for arrhythmias \cite{RAHHAL2016340,ACHARYA201781,DBLP:journals/corr/RajpurkarHHBN17}, suggesting that they may also be able to detect noise.

We begin with two models, a shallow 4-layer network and one with 6 layers, both with an input length of 784 samples at 300 Hz. Desiring greater precision in the results as well as the ability to consider the context of a given signal segment, we expand the input size to 10 seconds, generating a prediction for each 1-second segment. With this input size, we train CNNs with 6 and 16 layers and a residual network. We also try inputting 30 seconds of an ECG, still generating one prediction per second. Overall, the 16-layer 10-second model exhibits the best performance, demonstrating the ability to determine whether a given interval is noisier than, and thus would reduce the clarity of, the remainder of the signal.

\begin{figure}
\centering
\subfigure{\label{fig:annotation1}\includegraphics[width=\textwidth]{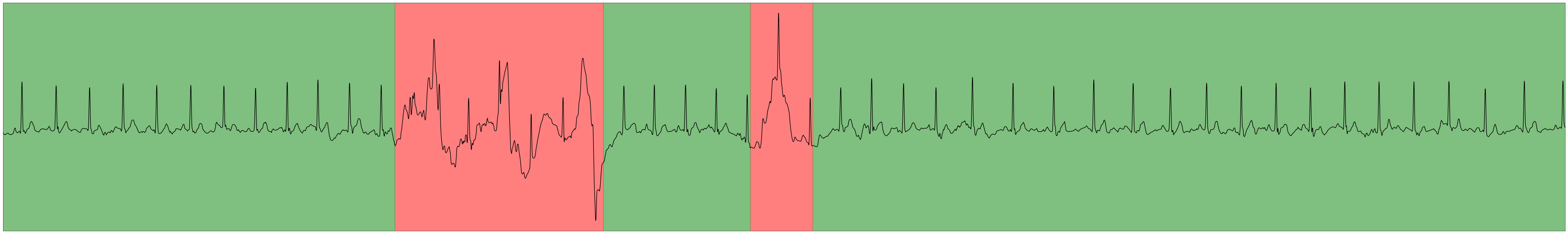}}
\subfigure{\label{fig:annotation2}\includegraphics[width=\textwidth]{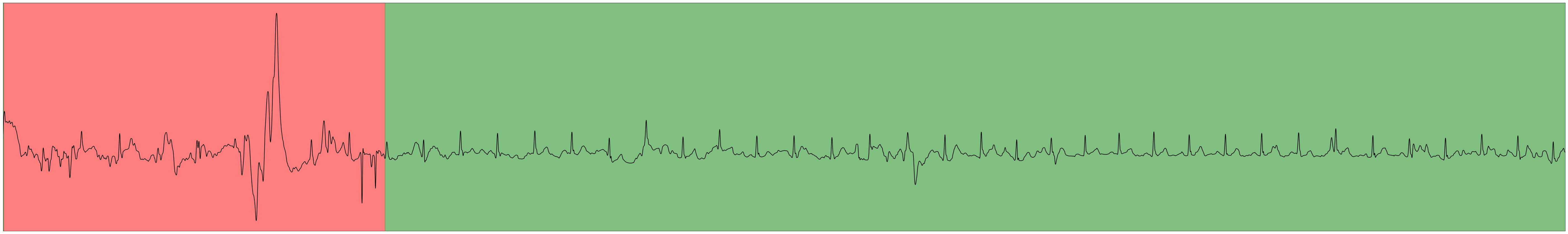}}
\caption{\label{fig:annotations}Example human-produced annotations.}
\end{figure}

\section{Dataset}
To our knowledge, no dataset of mobile ECG recordings labeled for noise exists. While the MIT-BIH Noise Stress Test Database does include noise in its ambulatory ECGs, it may not be sufficiently similar to the noise found in mobile ECGs because it is added artificially \cite{goldberger2000physiobank,moody1984noise,zhang2015ecg}.

Thus, we create a novel labeled dataset of 10,463 manually noise-annotated ECGs collected from AliveCor KardiaMobile devices, which are FDA-cleared and single-lead. A sampling frequency of 300 Hz was used. The majority of ECGs (56\%) are 30 seconds long, with an average length of 40.5 seconds. Depending on the input length of a given architecture, signals shorter than either 10 or 30 seconds were excluded. No distinction was made between different types or levels of noise while labeling. Significantly, to determine whether a given interval is noisy, we do not apply a universal noise threshold. Rather, we consider the degree of noise in the context of the overall signal, only labeling it if it is significantly noisier than the remainder of the signal. As a result, the same beat with an intermediate level of noise could be labeled as noise in a ECG that is otherwise almost entirely noise but not marked when the rest of the signal displays similar levels of noise. Given that our goal is to maximize the clarity of the features given to a model for diagnosis, this method makes the most sense because it will ultimately allow for the exclusion of noisier intervals. Note that although in signals with mixed quality we apply the "relative noise" heuristic, when an ECG has no readable intervals, we label the entire signal as noise. Conversely, for the many ECGs that are clean throughout, we do not mark any intervals of noise. Figure \ref{fig:annotations} gives an example ECG with annotated noise intervals.

When processing the data to create the train and test sets, we split the ECGs into either 784-sample (2.6-second) or 300-sample (1-second) segments. Majority voting is applied to determine whether a segment counts as noise or clean. As a result, segments for which 53\% and 100\% of their duration overlap with labeled noisy intervals will both be treated as entirely noise.

\section{Models}

\begin{wrapfigure}{R}{0.2\textwidth}
\centering
\includegraphics[width=0.3\textwidth]{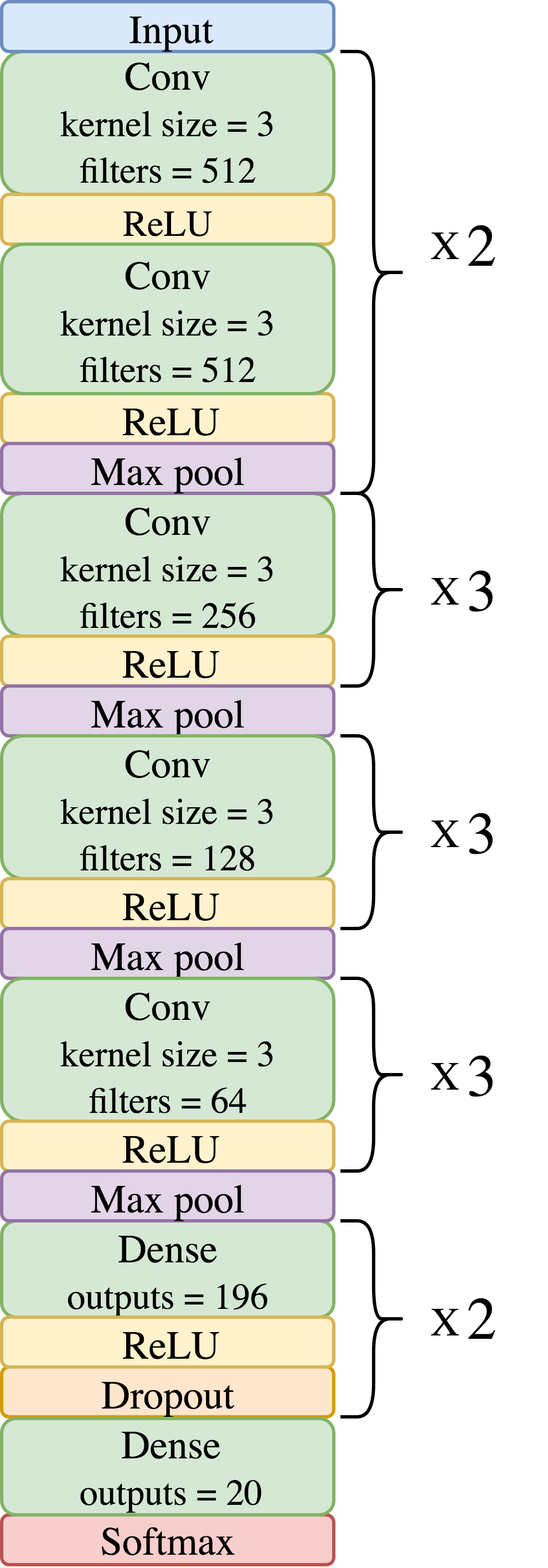}
\caption{\label{fig:abbreviatedVGG}Adapted VGG16 network with a signal length of 3000 (10 seconds). Diagram style inspired by \cite{DBLP:journals/corr/RajpurkarHHBN17}.}
\end{wrapfigure}

\subsection{Input Length: 784}
We began by implementing models with a signal length of 784 samples. With a sampling frequency of 300 Hz, this corresponds to approximately 2.6 seconds. The first model consisted of two one-dimensional convolutional layers followed by two fully-connected layers. The convolutional layers had 32 and 64 filters respectively and a kernel size of 8. The first fully-connected layer had 1024 units, while the second had 2 for binary classification. All layers except the final fully-connected layer were followed by ReLU activation. In one iteration, batch normalization layers were also added before each ReLU activation as a method of regularization \cite{ioffe2015batch}.

Suspecting that a deeper network would be necessary to fit the complexities of the data, we then implemented a model with 4 convolutional and 2 fully-connected layers. All convolutional layers had a kernel size of 8 and a stride length of 2. The number of filters were 32, 16, 8, and 4, respectively. The first fully-connected layer had 196 units, and the second again had 2. As before, we used ReLU activation after all layers but the last. No batch normalization was used in this model. We also tried inserting max pooling layers with a pool size of 4 and a stride length of 2 after each convolutional layer.

\subsection{Input Length: 3000}

Two issues revealed by the results of the 784-length inputs motivated the development of a model with an input length of 3000 samples generating 10 predictions each. First of all, a signal duration of 2.6 seconds did not provide sufficient precision. A high degree of variation in noise level can be present in the span of 2.6 seconds. Because the output was binary, the model was forced to classify the entire input as noisy or not. As a result, if just over half of the 2.6 seconds were noisy, it would not be possible to salvage the non-noisy interval. Because a small difference in noise levels between two segments could lead one to be labeled as noise and the other clean (for example, if one had 1.4 seconds of noise while the other had 1.2 seconds), the distinguishing features of the noise and clean examples were not always clear even to a human.

Second, without the context of the entire signal, the model could not determine whether a given input was noisier than its surrounding beats: it could only consider the 2.6 seconds in isolation. Consequently, the model could not satisfy the goal of determining whether a segment should be included in the final analysis of an ECG. Furthermore, because we used the “relative noise level” heuristic while creating the dataset, the definition of what constitutes noise varies between ECGs. 

To remedy these issues, we input 3000 samples, which would allow the model to account for the context of the signal. With a prediction every 300 samples, or 1 second, the model would also be able to better isolate noisy intervals.

As before, we experimented with several architectures. The first model was identical to the final 784-input model, with the modifications of a greater input size and 20 outputs in the last layer rather than 2. Next, we implemented a deeper model adapted from the VGG16 network \cite{simonyan2014very}, shown in Figure \ref{fig:abbreviatedVGG}. Because the final layer had 2 outputs instead of 1000 in the original VGG network, we inverted the sequence of filter sizes such that they decreased from 512 to 64 rather than increase. Finally, we tried a residual network with 6 convolutional layers \cite{he2016deep}.

\subsection{Input Length: 9000}

To provide the model with even more signal context to determine relative noise levels, we also tried inputting 30 seconds of an ECG signal, still generating predictions every second. The architecture of this model was the same as that of the first 10 second model we tried: 4 convolutional layers followed by 2 fully-connected.

\begin{table}
  \begin{center}
    \label{tab:table1}
    \begin{tabular}{r|llll}
      \textbf{Model} & \textbf{Sensitivity} & \textbf{Specificity} & \textbf{PPV} & \textbf{AUC}\\
      \hline
      \multicolumn{2}{@{}l}{\makecell{784-Length Models}}\\
      \midrule
      2 conv 2 FC & 0.865 & 0.954 & 0.293 & 0.976\\
      2 conv 2 FC + BN & 0.880 & 0.959 & 0.317 & 0.972\\
      4 conv 2 FC & 0.867 & 0.967 & 0.361 & 0.983\\
      4 conv 2 FC + pool & 0.817 & 0.978 & 0.450 & 0.984\\
      \midrule
       \multicolumn{2}{@{}l}{\makecell{3000-Length Models}}\\
      \midrule
      4 conv 2 FC + pool & 0.873 & 0.952 & 0.327 & 0.975\\
      VGG & 0.887 & 0.961 & 0.378 & 0.977\\
      ResNet & 0.757 & 0.984 & 0.558 & 0.966\\
      \midrule
      \multicolumn{2}{@{}l}{\makecell{9000-Length Model}}\\
      \midrule
      4 conv 2 FC + pool & 0.796 & 0.978 & 0.470 & 0.956\\
    \end{tabular}
    \caption{\label{tab:comparison}Comparison of model results.}
  \end{center}
\end{table}

\section{Results}
See Table \ref{tab:comparison} for a quantitative comparison of the models' performance. We evaluate them based on sensitivity, specificity, positive predictive value (PPV), and area under the receiver operating characteristic (ROC) curve (AUC).

\begin{figure}
\centering
\subfigure[Example of obvious noise; noise probability = 0.989.]{\label{fig:a}\includegraphics[width=60mm]{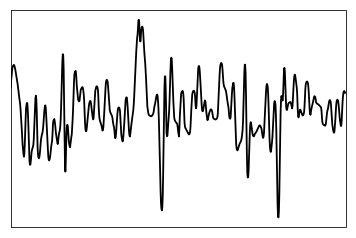}}
\subfigure[Example of significant noise lasting for less than half of the signal; noise probability = 0.4874.]{\label{fig:b}\includegraphics[width=60mm]{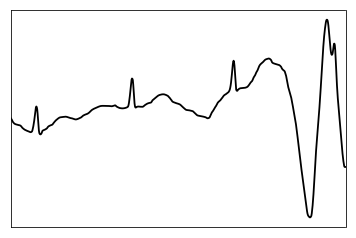}}
\caption{Example predictions given by the 6-layer 784-input network with pooling.}
\end{figure}

\subsection{Input Length: 784}
All of the models with 784-length inputs showed strong performance on the test set overall, with areas under the curve (AUCs) of 0.976 for the 4-layer network, 0.972 for the same network with batch normalization, 0.983 for the 6-layer network, and 0.984 for the 6-layer network with max pooling. Examples of extreme noise are easily detected, with predicted probabilities of noise often greater than 0.9; an example is shown in Figure \ref{fig:a}. Conversely, obviously clean examples frequently receive noise probabilities less than 0.001. Generally, the softmax probability appears to correlate well with the severity of noise. A critical issue is that with this set-up, the model is unable to distinguish between noisy and clean intervals within an input, which is problematic for signals that contain significant noise which spans less than half of the duration (see Figure \ref{fig:b} for an example). Due to the majority voting algorithm used in pre-processing, such signals are labeled as clean, even though ideally the noisy interval would be excluded.

\begin{figure}
\centering
\subfigure[Example prediction from 10-second VGG network.]{\label{fig:VGG}\includegraphics[width=\textwidth]{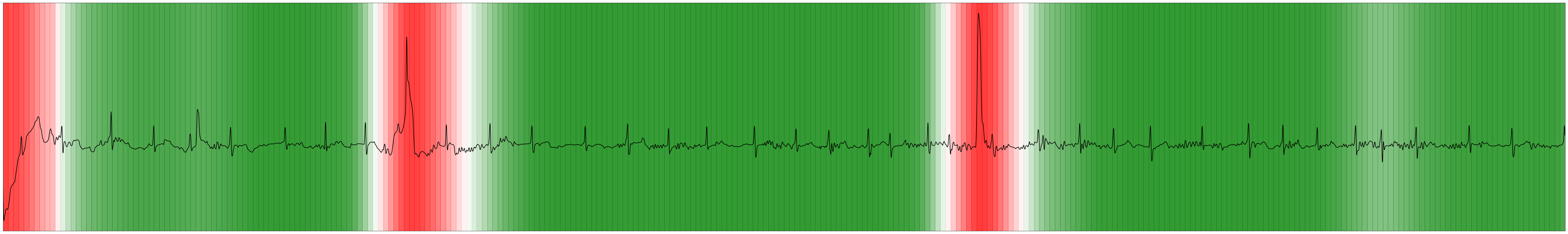}}
\subfigure[Example prediction from 30-second model.]{\label{fig:9000}\includegraphics[width=\textwidth]{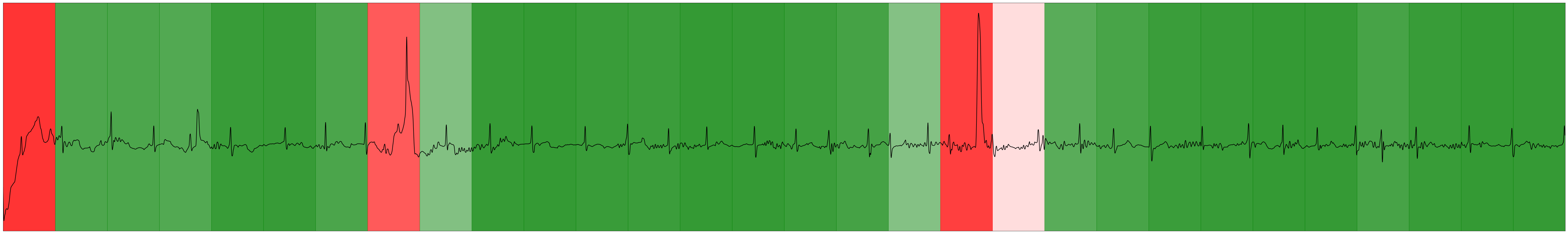}}\vspace{-0.5em}
\caption{}
\end{figure}

\subsection{Input Length: 3000}

When adapted to an input size of 3000, the 6-layer model had a slightly lower AUC of 0.975. However, it appears to perform better qualitatively than the 784-input model, demonstrating the ability to isolate intervals of noise and account for context more effectively. The VGG network had very similar performance to the 6-layer model, with a marginally improved AUC of 0.977. In contrast, the ResNet was slightly worse, with an AUC of 0.966.

Figure \ref{fig:VGG} shows an example prediction given by the VGG network. When producing the results, we stride over a 30-second ECG by 0.1 seconds, inputting 10 seconds at a time. The resulting probabilities are then aligned and averaged to increase accuracy and to be able to predict noise intervals beyond those with start and stop points that are multiples of 300. The color gradient represents the numerical value of the softmax output for the probability of noise, with values close to 1 translating to red, values close to 0 translating to green, and values near 0.5 translating to white. Additional results are given in the appendix.

\subsection{Input Length: 9000}

Although the 6-layer network with 30-second inputs still had a high AUC of 0.956, its performance was the worst out of all the architectures. In addition, because most of the signals are 30 seconds long, it is not possible to apply the same striding approach as used with 10-second inputs, which was essential for improving the precision of predicted noise intervals. A comparison of the results to the 10-second model is shown in Figure \ref{fig:9000}.

\section{Related Work}

Previous methods of noise detection in ECGs rely largely on raw statistical approaches. As one example, Chiarugi et al develop an algorithm to determine whether a given ECG channel should be included or excluded due to noise before attempting QRS detection. They calculate a noise index for each beat based on the ratio of the square average of the samples in the T-P interval and the region surrounding the R-peak. The beat is then given a rating of high, low, or intermediate noise based on the value of the noise index \cite{chiarugi2007adaptive}.

Lars Johannesen from the FDA proposes a rule-based algorithm to classify an entire ECG as noisy or clean. The algorithm considers the following factors: lead fail in all leads, global high-frequency noise, QRS detection, global low-frequency noise, and average beat quality. How each factor is assessed varies based on its characteristics. If the derivative is 0 across an entire signal for one lead, for example, it is classified as lead fail. To determine if an ECG possesses global high-frequency noise, the original signal is compared to that after a low-pass filter is applied. After each factor is assessed, an ECG is either discarded or continues to the next factor for assessment. ECGs that pass the final step are considered to be of acceptable quality \cite{johannesen2011assessment}.

Other approaches involve agglomerative clustering of portions of an ECG based on the extraction of features such as the number of times the signal crosses the baseline and the standard deviation \cite{rodrigues2017noise}, and the selection of explicit noise limits based on the requirements for the relevant task \cite{galeotti2013measurement}.

\section{Conclusions}

In this work, we show that convolutional neural networks can be applied to detect noise in ECGs recorded on mobile devices with strong performance nearing that of a human. Although models with an input of 2.6 seconds performed well on the test set, inputting 10 seconds produced the best results. This was due to the ability to both account for context and increase the granularity of predicted noise intervals. Out of the architectures we implemented, a modified VGG16 network showed the best performance, with an AUC of 0.977.

One limitation of our work that suggests a direction for future research is the inconsistency in the labels of the dataset. Since whether an interval should be labeled as noise or not is subjective, it is possible that labels on different ECGs could contradict each other. In addition, because we use a binary definition of noise, a slight difference in noise severity in two different segments could lead one to be labeled as noise and the other clean. This approach fails to account for the spectrum of noise levels present in ECGs. However, the observation that the predicted probabilities of noise that our model generates correlate well with the level of noise suggests that this is not a major issue. Still, it would be worth trying to classify noise into different categories (for example, no noise, little noise, intermediate noise, and very noisy) or rating the noise level on a numerical scale to determine if performance is improved. Another potential modification to our labeling approach would be to specify different types of noise (radio frequency, motion, electromyogram, etc). Then, algorithms targeted to reduce specific noise types could be applied.

Furthermore, there is also the possibility that recurrent neural networks (RNNs) could perform well on this task due to their specialization for sequential inputs, including time series. The forecasted continued time series predicted by an RNN could be compared to the actual sequence, with a significant difference indicating noise.

At a time when cardiovascular disease poses a dire threat to the world’s population, detecting intervals of noise in ECGs will improve the performance of existing arrhythmia classification algorithms, increasing the power of mobile ECG-recording devices as a means of monitoring heart health. 

\section{\ackname}{
The first author's work was supported by a summer internship at AliveCor, Inc.
}

\bibliographystyle{plain}
\bibliography{sample}

\appendix
\section{Additional results from VGG network}
\begin{figure}[H]
\subfigure{\label{fig:aVGG}\includegraphics[width=\textwidth]{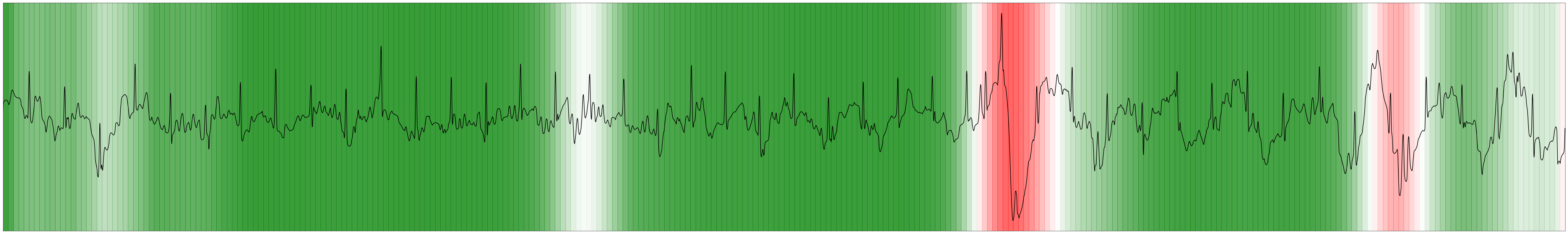}}
\subfigure{\label{fig:bVGG}\includegraphics[width=\textwidth]{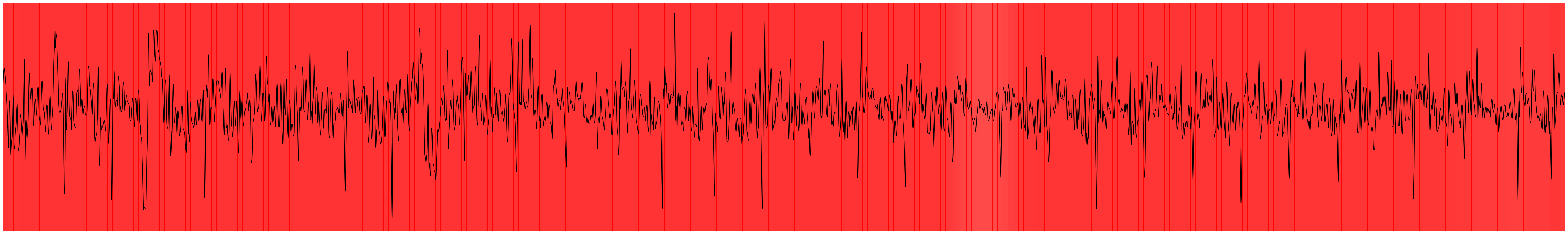}}
\subfigure{\label{fig:cVGG}\includegraphics[width=\textwidth]{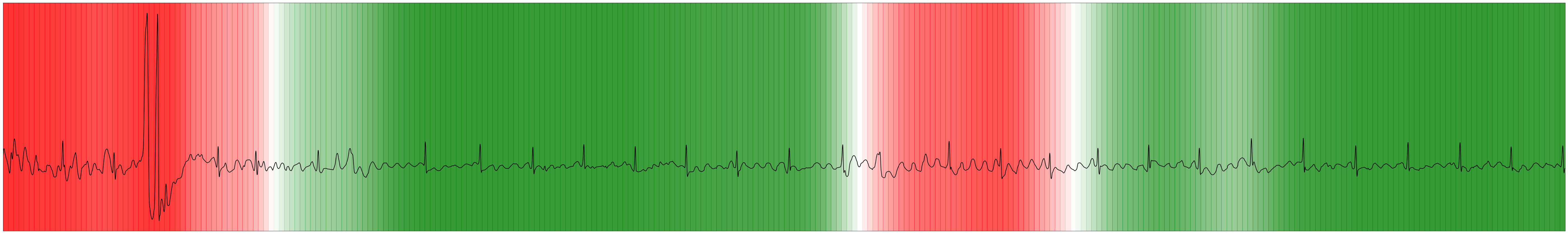}}
\subfigure{\label{fig:dVGG}\includegraphics[width=\textwidth]{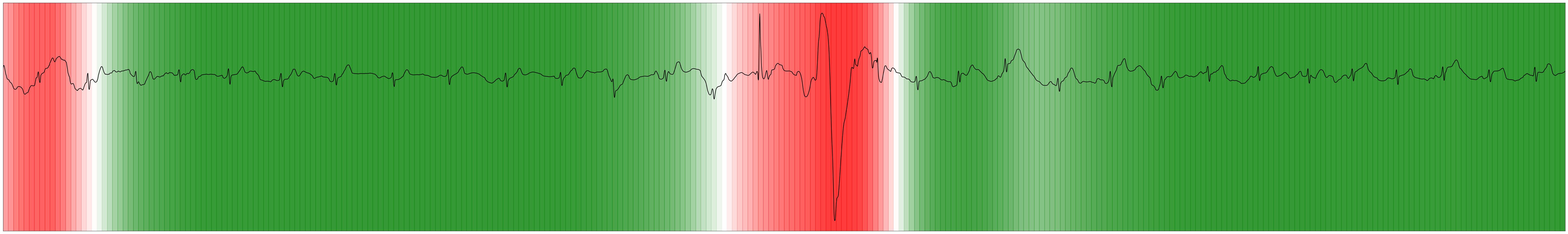}}
\subfigure{\label{fig:eVGG}\includegraphics[width=\textwidth]{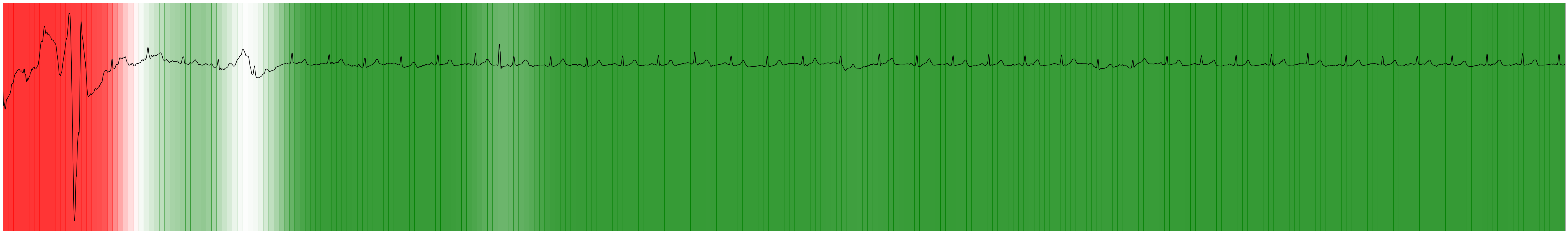}}
\subfigure{\label{fig:fVGG}\includegraphics[width=\textwidth]{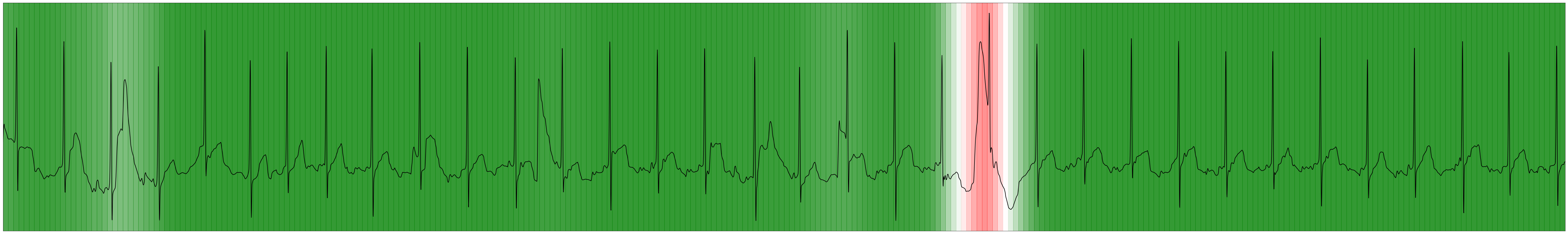}}
\subfigure{\label{fig:gVGG}\includegraphics[width=\textwidth]{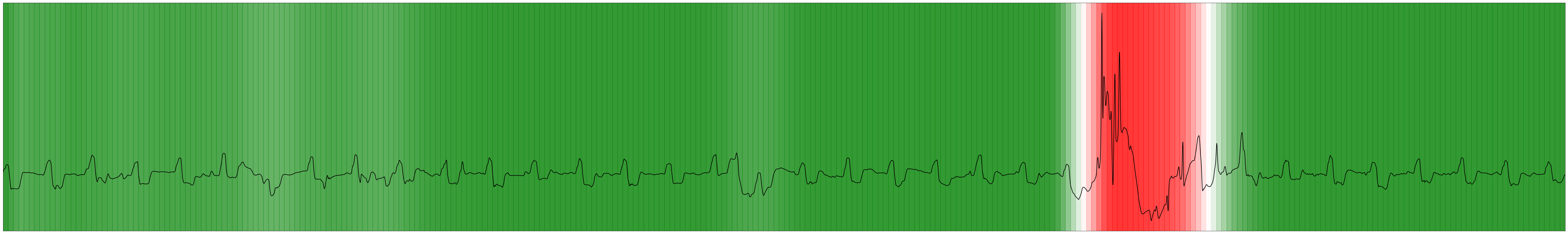}}
\subfigure{\label{fig:hVGG}\includegraphics[width=\textwidth]{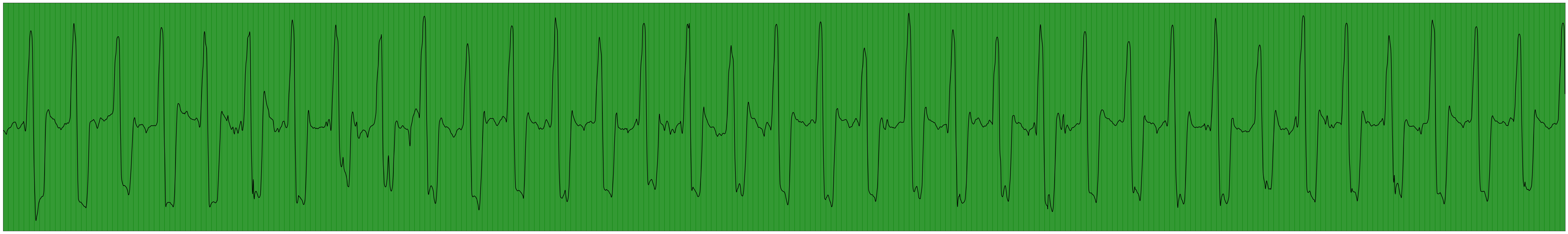}}
\subfigure{\label{fig:iVGG}\includegraphics[width=\textwidth]{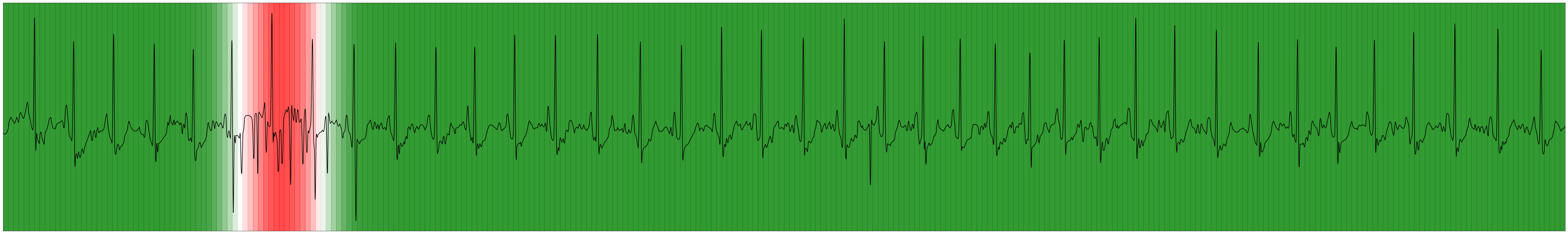}}

\end{figure}

\end{document}